\documentclass[twocolumn, nofootinbib, secnumarabic, amssymb, preprintnumbers, superscriptaddress, aps, prd]{revtex4-1}

\usepackage{graphicx}

\usepackage{cancel,tabularx,moreverb,fancybox,amsmath,float,bm,braket,slashbox,txfonts,amssymb,bm,accents}
\usepackage{latexsym}
\usepackage{here}
\usepackage{amsfonts}
\usepackage[normalem]{ulem}
\usepackage{color}
\usepackage{bbold}
\usepackage{ulem}
\usepackage{graphicx}
\usepackage{hyperref}

\newcommand{\pa}[1]{\left(#1 \right)}

\def\be{\begin{equation}}
\def\ee{\end{equation}}
\def\ba{\begin{eqnarray}}
\def\ea{\end{eqnarray}}

\def\tr{{\text{tr}}}

\begin{document}

\preprint{}
\title{Islands for Reflected Entropy}
\date{\today}
\author{Venkatesa Chandrasekaran}\email[ven\_chandrasekaran@berkeley.edu]{}
\author{Masamichi Miyaji}\email[masamichimiyaji@berkeley.edu]{}
\author{Pratik Rath}\email[pratik\_rath@berkeley.edu]{}
\affiliation{\it Berkeley Center for Theoretical Physics, Department of Physics, University of California, Berkeley, CA 94720, USA}

\begin{abstract} 
Recent work has demonstrated the need to include contributions from {\it entanglement islands} when computing the entanglement entropy in QFT states coupled to regions of semiclassical gravity.
We propose a new formula for the reflected entropy that includes additional contributions from such islands.
We derive this formula from the gravitational path integral by finding additional saddles that include generalized replica wormholes.
We also demonstrate that our covariant formula satisfies all the inequalities required of the reflected entropy.
We use this formula in various examples that demonstrate its relevance in illustrating the structure of multipartite entanglement that are invisible to the entropies.

\end{abstract}

\maketitle
\noindent

\section{Introduction}
The black hole information paradox, in its various versions, has been a longstanding hurdle in our understanding of quantum gravity \cite{Hawking:1974rv,Hawking:1974sw, Mathur:2009hf,Almheiri:2012rt,Almheiri:2013hfa}. 
It was long believed that a UV complete description of the black hole evaporation process would be a necessary ingredient for resolving the paradox.
However, significant progress has been made recently in resolving this issue within the realm of semiclassical gravity by invoking a new rule - the so called  ``islands formula" \cite{Penington:2019npb, Almheiri:2019psf, Almheiri:2019hni, Penington:2019kki, Almheiri:2019qdq} \footnote{For other related work, see \cite{Almheiri:2020cfm} and references therein}.

The proposal to compute the fine-grained von Neumann entropy $S(A)$ of the subregion $A$ is given by
\begin{align}\label{eq:island}
S(A)&= S^{(\text{eff})}(A\cup \text{Is}(A))+\frac{\text{Area}[\partial \text{Is}(A)]}{4G_{N}},
\end{align}
where $S^{(\text{eff})}(A\cup \text{Is}(A))$ represents the von Neumann entropy computed in the effective semiclassical theory that includes contributions from possible {\it entanglement islands}, in the gravitating region, denoted $\text{Is}(A)$.
In a general time-dependent situation, the location of the island is computed using a maximin procedure as we review in Section~\ref{sec:defn} \cite{Akers:2019lzs}. 
This formula was first justified by considering holographic matter which itself has a bulk dual in \cite{Almheiri:2019hni}, and then proved using the gravitational path integral in \cite{Penington:2019kki,Almheiri:2019qdq}.
An essential feature of the proof was the inclusion of new saddles in the path integral that are wormhole solutions connecting different replica manifolds. 
Importantly, entanglement islands can also be understood as a part of the {\it entanglement wedge} \cite{Czech:2012bh, Dong:2016eik, Cotler:2017erl, Kusuki:2019hcg}, which is determined by Ryu-Takayanagi surface\cite{Ryu:2006bv, Ryu:2006ef, Hubeny:2007xt}, within which bulk operators can be reconstructed by an observer sitting at the AdS boundary. For discussions on such reconstruction in the context of black hole information, see \cite{Penington:2019kki, Nomura:2019qps, Nomura:2019dlz, Brown:2019rox}.  

In this paper we consider a bipartite correlation measure, the {\it reflected entropy} \cite{Dutta:2019gen}. 
Given a reduced density matrix $\rho_{AB}$, one can canonically purify the state as $\ket{\sqrt{\rho_{AB}}}_{ABA'B'}$ in a doubled copy of the Hilbert space which includes subregions $A'$ and $B'$.
This is the familiar procedure that one considers in going from the thermal density matrix to the thermofield double state \cite{Maldacena:2001kr}.
Given this state, the reflected entropy is defined as
\begin{align}\label{eq:SR}
    S_R(A:B)&=S(AA').
\end{align}
The reflected entropy serves as a measure of correlations between subregions $A$ and $B$, which includes both classical and quantum correlations \cite{Umemoto:2019jlz,Levin:2019krg}.
It satisfies various inequalities that make it consistent with this interpretation.
This quantity, being a different measure of multipartite entanglement, often allows us to distinguish the fine structure of entanglement which the von Neumann entropy cannot capture \cite{Akers:2019gcv}.
Interestingly, it has a simple holographic dual, the ``entanglement wedge cross section" which was originally proposed as a dual to the entanglement of purification, $E_P(A:B)$ \cite{Takayanagi:2017knl,Nguyen:2017yqw} \footnote{Note that $E_P(A:B)$ is a far harder quantity to compute away from Gaussian approximations \cite{Bhattacharyya:2018sbw, Bhattacharyya:2019tsi} and hence, it is difficult to prove statements about it. 
However, it is quite plausible that our modified formula for the reflected entropy $S_R(A:B)$ might extend to the case of $E_P(A:B)$. See \cite{Caputa:2018xuf, Umemoto:2019jlz}}.

In this paper, we propose an ``islands formula" that captures reflected entropy in the presence of entanglement islands by relating it to the reflected entropy in the effective semiclassical theory.
Concretely, the proposal is
\begin{multline} \label{eq:islandSR}
S_R(A:B)=S_R^{(\text{eff})}(A\cup \text{Is}_R(A):B\cup \text{Is}_R(B))\\
+\frac{\text{Area}[\partial \text{Is}_R(A)\cap \partial \text{Is}_R(B)]}{2G_{N}},
\end{multline}
where the {\it reflected entropy islands}, denoted $\text{Is}_R(A)$ and $\text{Is}_R(B)$, split the entanglement island $\text{Is}(AB)$ into two parts.
We emphasize that these are in general different from the entanglement islands $\text{Is}(A)$ and $\text{Is}(B)$, and are also computed by a maximin procedure that we describe in Section~\ref{sec:defn}.

\begin{figure}[h]

  \includegraphics[clip,width=0.8\columnwidth]{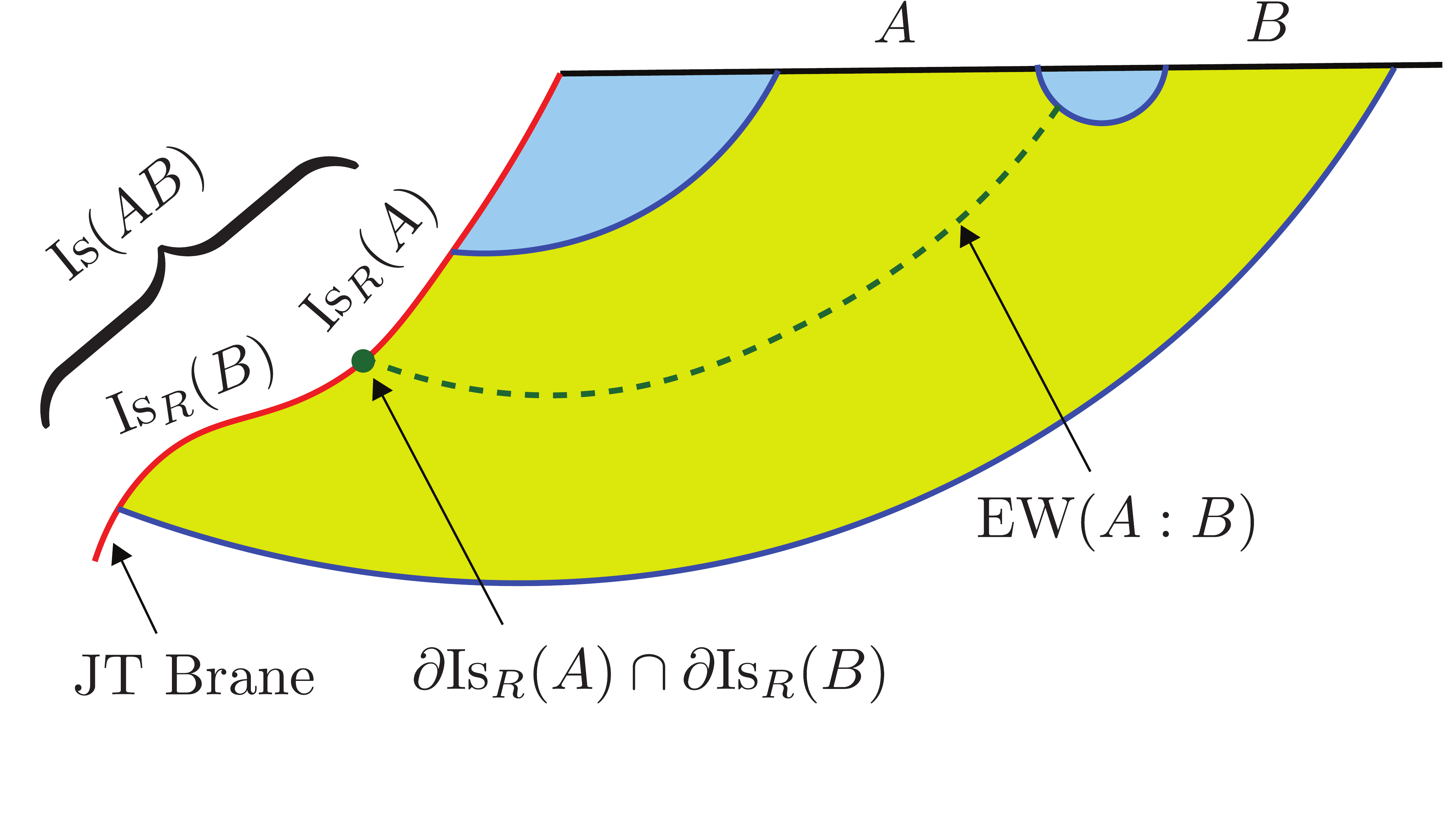}

\caption{A $d$ dimensional BCFT has a $d$ dimensional effective description in terms of a gravitating brane coupled to flat space.
In the presence of holographic matter, this effective theory itself has a $d+1$ dimensional bulk dual.
The reflected entropy of the regions $A$ and $B$ in the BCFT can be computed using the entanglement wedge cross section $\text{EW}(A:B)$ in the $d+1$ dimensional bulk dual.
From the perspective of the effective $d$ dimensional theory, this leads to the islands formula of Eqn.~(\ref{eq:islandSR}).}
\label{fig:EWCS}
\end{figure}

This formula can be motivated by considering a $d$ dimensional BCFT dual to $d$ dimensional holographic matter, which itself has a $d+1$ dimensional bulk dual, i.e., the so called double holography scenario of \cite{Almheiri:2019hni}.
In this case, we can simply use the entanglement wedge cross section to compute the reflected entropy.
As seen in Figure~\ref{fig:EWCS}, the $d+1$ dimensional entanglement wedge cross section could reach the island, in which case there would be additional contributions from the perspective of the effective $d$ dimensional theory.
This can be captured by the modified formula in Eqn.~(\ref{eq:islandSR}).

In the remainder of this paper, we try to demonstrate how the proposed formula is applicable to much more general situations.
In Section~\ref{sec:defn}, we describe the proposed formula in Eqn.~(\ref{eq:islandSR}) in detail, utilizing a maximin procedure to demonstrate that it indeed satisfies all the required properties of reflected entropy in a general time-dependent situation. 
In Section~\ref{sec:derive}, we then provide a derivation of the above formula with a replica trick argument using the Euclidean gravitational path integral.
We restrict to time-independent situations for convenience but the arguments are general enough that they can be extended to time-dependent cases.
In Section~\ref{sec:examples}, we then illustrate the use of the islands formula in various examples.
This analysis sheds light on the structure of multipartite entanglement in the Hawking radiation, and demonstrates an interesting phase transition in the behaviour that isn't captured by the entanglement entropy.
Finally, we discuss some implications of our analysis and interesting future directions in Section~\ref{sec:disc}.

Note: This paper is being released in coordination with \cite{other} which has some overlap with our work.




\section{Islands Formula}\label{sec:defn}
\subsection{Proposal}

In the presence of gravitating regions, the maximin formula for computing entanglement entropy is given by \cite{Akers:2019lzs}
\begin{align}\label{eq:island2}
S(A)&=\underset{\Sigma}{\text{Max}}\underset{I\subset \Sigma}{\text{Min}}\Big[S^{(\text{eff})}(A\cup I)+\frac{\text{Area}[\partial I]}{4G_{N}}\Big],
\end{align}
where one considers arbitrary Cauchy slices $\Sigma$ that include $A$ and finds the island $I$ on each $\Sigma$ that minimizes the hybrid entropy functional in Eqn.~(\ref{eq:island2}) \footnote{In the presence of appropriate boundary conditions that allow for unitary evolution in the boundary, we can relax this condition to include only $\partial A$.}.
Then, we finally maximize over the choice of slice which results in an entanglement island that extremizes the hybrid entropy, which we shall denote as $\text{Is}(A)$.

In analogy with the holographic proposal in \cite{Dutta:2019gen}, we propose to use a similar procedure for computing the reflected entropy between regions $A$ and $B$.
One first fixes the entanglement wedge of the region $AB$ by finding the relevant island $\text{Is}(AB)$.
Then, we split it into two portions divided by a ``minimal" entanglement wedge cross section on arbitrary Cauchy slices $\Sigma$ that contain $AB$ and $\text{Is}(AB)$.
More precisely, we choose $I_A$ and $I_B$ such that $I_A\cup I_B=D(\text{Is}(AB)) \cap \Sigma$ where $D(\text{Is}(AB))$ is the causal domain of dependence of $\text{Is}(A)$.
Finally one maximizes over the choice of slice to obtain
\begin{multline}\label{eq:islandSR2}
S_R(A:B)=\underset{\Sigma}{\text{Max}}\underset{I_A\cup I_B\subset \Sigma}{\text{Min}}\Big[S_R^{(\text{eff})}(A\cup I_A:B\cup I_B)+\\
\frac{\text{Area}[\partial I_A\cap \partial I_B]}{2G_{N}}\Big].
\end{multline}

It is useful to think of Eqn.~(\ref{eq:islandSR2}) as applying the usual islands formula, Eqn.~(\ref{eq:island2}), in the state $\ket{\sqrt{\rho_{AB}}}$ \footnote{We thank Thomas Faulkner for discussions about this.}.
In the non-gravitating regions, it is clear what this operation does, but it is much less clear how the gravitating regions change under this procedure.
It has been proposed that one should glue a CPT conjugate copy of the geometry across the quantum extremal surface for the subregion $AB$ \cite{Engelhardt:2017aux,Engelhardt:2018kcs,Dutta:2019gen,Bousso:2019dxk}.
The bulk state is chosen to be the canonical purification of the original bulk state.
Applying Eqn.~(\ref{eq:island2}) to this state leads us to Eqn.~(\ref{eq:islandSR2}).
We conjecture that just like the entanglement islands formula, our reflected entropy islands formula works to all orders in $G_N$ perturbation theory \cite{Dong:2017xht}.
There are various subtleties with graviton entanglement entropy that need to be resolved to make this concrete, but at the least we expect it to work to $\mathcal{O}(1)$, i.e., first subleading order.

\subsection{Consistency Checks}

Having formulated our conjecture, we now show that our proposed formula satisfies the same properties as the reflected entropy $S_R(A:B)$.
The entanglement of purification also satisfies a similar set of inequalities, and hence, the evidence provided in this section would suggest the same formula works for $E_P(A:B)$ as well.
In fact, there are an additional set of inequalities relevant to $E_P(A:B)$ that our formula also satisfies as we show in Appendix~\ref{app:ineq}.
These properties essentially follow from the definition in Eqn.~(\ref{eq:islandSR2}) and from the relative locations of entanglement islands, as dictated by the quantum focussing conjecture which we shall assume for this purpose \cite{Bousso:2015mna}.

In the following, we will use the notation $S(A,A')_{\Sigma}=S^{(\text{eff})}(AA')+\frac{\text{Area}[\partial A']}{4G_{N}}$ for the hybrid entropy on $\Sigma$ with $A'$ chosen as the candidate island before extremizing.
Similarly, we define $S_R(A,A':B,B')_{\Sigma}=S_R^{(\text{eff})}(AA':BB')+\frac{\text{Area}[\partial(A'\cap B')]}{2G_{N}}$ for the reflected entropy, where the quantity is not yet extremized. 

\subsubsection*{Properties}

\begin{itemize}
    \item $S_R(A:B)\geq 0$ 
 \end{itemize}
 
    This is obvious from the definition since both $S_R^{(\text{eff})}$ and $\text{Area}$ are non-negative quantities.
  \begin{itemize}
    \item $S_R(A:B)$ is invariant upon acting with local unitaries on regions $A$ and $B$ respectively.
    \end{itemize}
    To show this, it is useful to think about the canonically purified state $\ket{\sqrt{\rho_{AB}}}$.
    In this state, the entanglement wedge of subregion $A$ must lie inside the entanglement wedge of $A\cup A^*$, where $A^*$ is the CPT conjugate subregion to $A$.
    This property, termed entanglement wedge nesting, has been proven in \cite{Akers:2019lzs}.
    This implies that $D(\text{Is}(A))\subset D(\text{Is}_R(A))$.
    A local unitary acting on subregion $A$ in the UV theory, acts as a local unitary on the region $A\cup \text{Is}(A)$ in the effective theory.
    Thus, the area of $\partial \text{Is}_R(A)$ and hence, $\partial \text{Is}_R(A) \cap \partial \text{Is}_R(B)$ cannot be affected.
    Further, the bulk state in the region $A\cup\text{Is}_R(A)$ is modified by a local unitary, and thus, $S_R^{(\text{eff})}$ remains invariant.
    The same argument above can be repeated for local unitaries acting on subregion $B$.
    
\begin{itemize}
    \item $S_R(A:B)=2S(A)=2S(B)$ when $\rho_{AB}$ is pure.
\end{itemize}

Purity of $\rho_{AB}$ implies that $AB\cup\text{Is}(AB)$ is a complete Cauchy slice of the effective theory.
In particular, the bulk effective field theory state on region $AB\cup\text{Is}(AB)$ is pure and hence, $S_R^{(\text{eff})}(A\cup I_A:B\cup I_B)=2S(A\cup I_A)=2S(B\cup I_B)$ for any choice of candidate reflected entropy islands $I_A$ and $I_B$.
Further, purity implies that $I_A\cup I_B=\text{Is}(AB)$ spans the entire gravitating region, and $\partial(\text{Is}(A)\cup\text{Is}(B))=\varnothing$.
Thus, $\partial I_A=\partial I_B$ and using the islands formula in Eqn.~(\ref{eq:islandSR2}), we see that the optimization is identical to that in Eqn.~(\ref{eq:island2}).
Thus, $\text{Is}(A)=\text{Is}_R(A)$ and we find the above equality.

\begin{itemize}
    \item $S_R(A:B)\leq 2\text{Min}(S(A),~S(B))$.
\end{itemize}

Let $\Sigma_{A:B}$ denote the Cauchy slice on which the $S_R(A:B)$ optimization in Eqn.~(\ref{eq:islandSR2}) is maximized.
Consider the subregion $A''$ of $\Sigma_{A:B}$ that minimizes $S(A,A'')_{\Sigma_{A:B}}$, i.e., the candidate minimal entanglement island on $\Sigma_{A:B}$. 
Now, since $A''\subset \text{Is}(AB)$ using nesting, we can define $B''=\text{Is}(AB)\setminus A''$ \cite{Wall:2012uf,Akers:2019lzs}.
Then, from the maximin procedure, we have
\begin{align*}
S_R(A:B)&\leq S_R(A,A'':B,B'')_{\Sigma_{A:B}} &\text{(minimization)}\,\\
&\leq 2 S(A,A'')_{\Sigma_{A:B}} &\text{(property of $S_R^{(\text{eff})}$)}\,\\
&\leq 2 S(A)  &\text{(maximization)}.
\end{align*}

A similar argument works for subregion $B$ and thus, we obtain the above inequality.

\begin{itemize}
    \item $S_R(A:B)\geq I(A:B)$.\label{mutual}
\end{itemize}

Consider the Cauchy slice $\Sigma$, on which each of $A\cup \text{Is}(A)$, $B\cup\text{Is}(B)$ and $AB\cup\text{Is}(AB)$ are extremal, which exists by the nesting arguments made in \cite{Wall:2012uf,Akers:2019lzs}.
$\Sigma \cap D(\text{ls}(AB))$, i.e., the island portion of the Cauchy slice, can be partitioned into $A'$ and $B'$, which minimizes $S_R(A,A':B,B')_{\Sigma}$. 
In general, we have $\text{Is}(A)\cup\text{Is}(B)\subset A'\cup B'$. 
Then, from the maximin procedure, we obtain
\begin{align*}
I(A:B)&= S(A,\text{Is}(A))_{\Sigma}+S(B,\text{Is}(B))_{\Sigma}-S(AB,\text{Is}(AB))_{\Sigma}\nonumber\\
&\leq S(A,A')_{\Sigma}+S(B,B')_{\Sigma}-S(AB,A'B')_{\Sigma}\nonumber\\
&= I^{(\text{eff})}(AA':BB')+\frac{\text{Area}[\partial A'\cap \partial B']}{2G_{N}}\nonumber\\
&\leq S_R(AA':BB')_{\Sigma} \leq S_R(A:B),
\end{align*}
where the second line follows from minimization and the last line follows from properties of $S_R^{(eff)}$ and maximization.

\section{Path integral Argument}\label{sec:derive}

\subsection*{Argument from Islands Formula}

In this section, we give an argument for our proposed islands formula using the gravitational path integral.
This argument is quite similar to that made in \cite{Dutta:2019gen} with the added ingredient being the presence of entanglement islands in the bulk.
We first describe the holographic dual of the canonically purified state $\ket{\sqrt{\rho_{AB}}}$.
Using the islands formula for entanglement entropy in the dual spacetime then gives us our proposed islands formula.
In what follows, we will use large central charge $c$ to ignore subtleties coming from gravitational fluctuations, however, it is quite plausible that these issues may be resolved without requiring large $c$ \cite{Dong:2017xht}.

\begin{figure}[h]

  \includegraphics[clip,width=0.7\columnwidth]{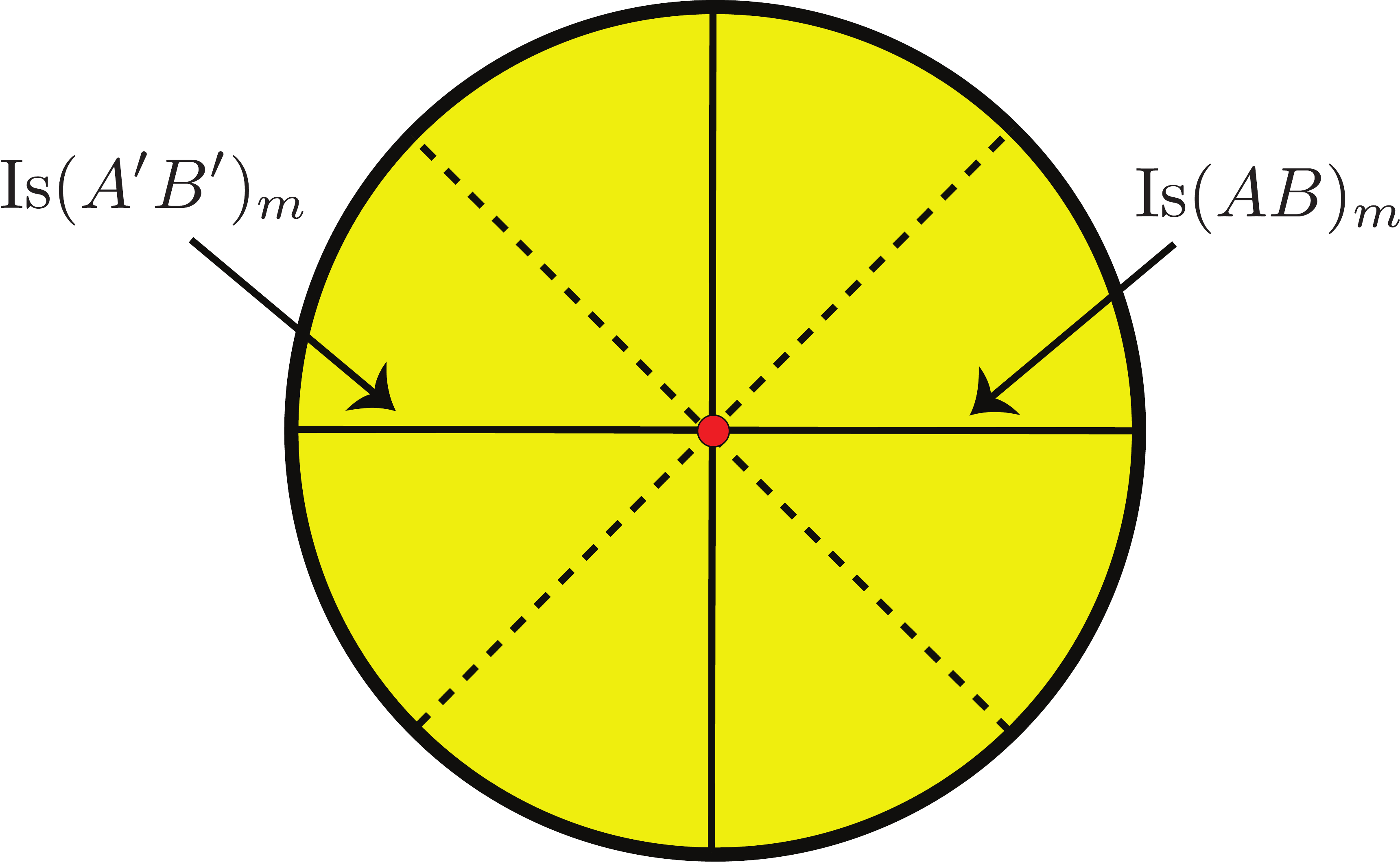}

\caption{The gravitational region $\mathcal{M}^{\text{bulk}}_{m}$ (shaded yellow) of the manifold $\mathcal{M}_{m}$ that computes $Z_m$ for $m=4$ is depicted here. In addition to a cyclic $\mathbb{Z}_m$ symmetry, we have a $\mathbb{Z}_2$ reflection symmetry which allows us to consider the bulk dual to the state $\ket{\rho_{AB}^{m/2}}$ by cutting open the path integral in half about the horizontal axis $\Sigma_m$. The Cauchy slice $\Sigma_m$ is made up of two pieces, that are denoted $\text{Is}(AB)_m$ and $\text{Is}(A'B')_m$, which become the entanglement islands of the respective regions in the limit $m\rightarrow1$. The red dot denotes the fixed point of $\mathbb{Z}_m$ symmetry that becomes the quantum extremal surface as $m\rightarrow1$. The dashed lines represent the complementary region to the island which has been traced out.}
\label{fig:GNS}
\end{figure}

In order to understand the construction, we first consider the path integral that computes $Z_m=\tr\,\rho_{AB}^m$ for even integer $m$.
The path integral is evaluated on a manifold $\mathcal{M}_{m}$ which includes a non-gravitational portion $\mathcal{M}^{\text{fixed}}_{m}$ with fixed geometry, as well as gravitational regions with dynamical geometry denoted $\mathcal{M}^{\text{bulk}}_{m}$.
In the gravitational region, we only specify asymptotic boundary conditions as fixed by $\mathcal{M}^{\text{fixed}}_{m}$ and integrate over all possible geometries consistent with them.
Further, we use the saddle point approximation as $G_N\rightarrow0$, and we expect a single geometry $\mathcal{M}^{\text{bulk}}_{m}$ to dominate the result.
Thus, we obtain the replicated manifold $\mathcal{M}_{m}=\mathcal{M}^{\text{fixed}}_{m}\cup\mathcal{M}^{\text{bulk}}_{m}$

As shown in Figure~\ref{fig:GNS}, we assume that the manifold $\mathcal{M}_{m}$ that dominates the path integral has a $\mathbb{Z}_m$ symmetry that cycles the individual replica copies in $\mathcal{M}^{\text{fixed}}_{m}$ and extends into $\mathcal{M}^{\text{bulk}}_{m}$.
In order to construct the state $\ket{\sqrt{\rho_{AB}}}$, we need to analytically continue $Z_m$ in the parameter $m$.
The prescription for analytic continuation is the one originally described in \cite{Lewkowycz:2013nqa}.

Using the $\mathbb{Z}_m$ symmetry, we consider an orbifold geometry $\tilde{\mathcal{M}}_m = \mathcal{M}_m/\mathbb{Z}_m$.
$\tilde{\mathcal{M}}_m$ is a geometry with a conical defect with opening angle $2\pi/m$ that comes from the fixed point of $\mathbb{Z}_m$ symmetry in $\mathcal{M}^{\text{bulk}}_m$.
Simultaneously, the orbifold in the non-gravitational region induces twist operators at the entangling surface on the original manifold $\mathcal{M}^{\text{fixed}}_1$.
Now for arbitrary $m$, one can simply define $\tilde{\mathcal{M}}^{\text{bulk}}_m$ by the bulk saddle with a conical defect of opening angle $2\pi/m$ that solves Einstein's equations everywhere else and asymptotically satisfies the correct boundary conditions prescribed by $\mathcal{M}^{\text{fixed}}_1$.

In addition, we assume that $\mathcal{M}_{m}$ has a $\mathbb{Z}_2$ symmetry owing to its time reversal invariance as seen in Figure~\ref{fig:GNS}.
This allows us to cut open the path integral at a $\mathbb{Z}_2$ symmetric Cauchy slice $\Sigma_m$ to construct the leading approximation to the bulk region corresponding to the fine-grained state $\ket{\rho_{AB}^{m/2}}$.
$\Sigma_m$ then provides an initial data surface which can be evolved to find the entire Lorentzian spacetime.
Having constructed the spacetime, we can now apply the entanglement islands formula, Eqn.~(\ref{eq:island}), for computing $S(AA')$, i.e., the reflected entropy, $S_R(A:B)$.
Due to time reversal invariance, the quantum extremal surface lies on $\Sigma_m$ and gives an island-like contribution to the reflected entropy.

Finally, we can take the limit $m\rightarrow1$, which gives us a spacetime as seen in Figure~\ref{fig:sigma} where $\Sigma_1$ is made up of two copies of the island region $\text{Is}(AB)$ glued together at the quantum extremal surface for the subregion $AB$, i.e., $\partial \text{Is}(AB)$.
This is identical to the construction proposed in \cite{Engelhardt:2018kcs} as the holographic dual of canonical purification.
Importantly, the island appears as an additional closed universe portion of the bulk which is entangled with the region $AB\cup A'B'$.
Now using the formula in Eqn.~(\ref{eq:island}) one obtains our proposed formula Eqn.~(\ref{eq:islandSR}).\footnote{Note that it was important that here we computed the entanglement entropy at arbitrary $m$ before continuing to $m=1$.
This specific order of limits was discussed in \cite{Dutta:2019gen,Kusuki:2019evw} and seems to result in a sensible analytic continuation that commutes with the $G_N\rightarrow0$ limit.
Aspects of this analytic continuation will be discussed in \cite{pagereflected}.
}

\subsection*{Heuristic Argument from Replica Trick}

\begin{figure}[h]

  \includegraphics[clip,width=\columnwidth]{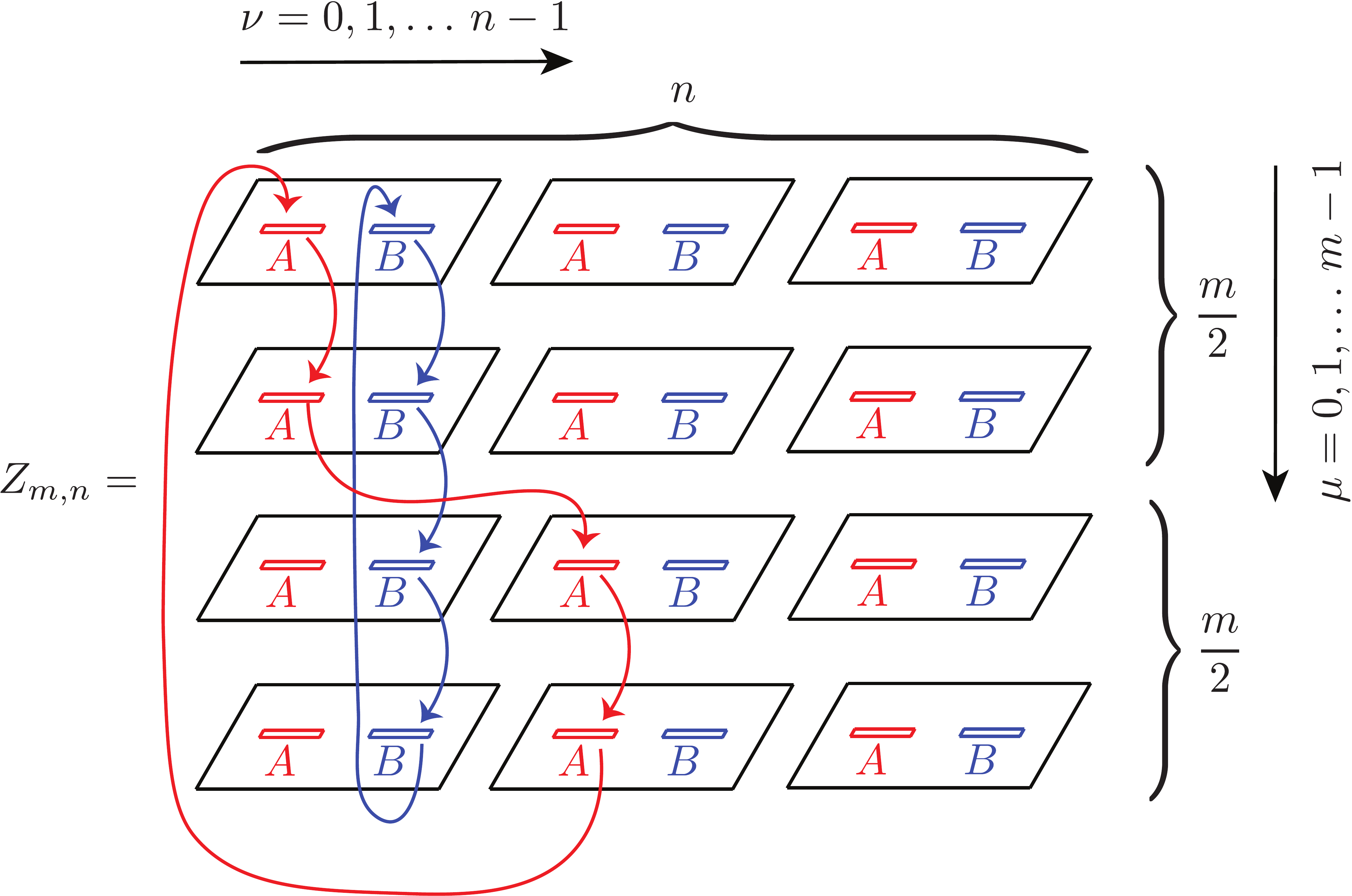}

\caption{The manifold $\mathcal{M}_{m,n}$ involves gluing the subregions B cyclically in the vertical $\mu$ direction, whereas the subregions A are glued together cyclically in the vertical direction upto a cyclic twist, in the horizontal $\nu$ direction, at $\mu=0,\frac{m}{2}$.
}
\label{fig:replica}
\end{figure}

An alternate way to perform the calculation is to consider a more complicated replica trick \cite{Dutta:2019gen}.
In order to compute the reflected entropy, we can first compute
\begin{align}\label{eq:Zmn}
    Z_{m,n}&=\tr_{AA'}\Bigg(\tr_{BB'}\ket{\rho_{AB}^{m/2}}\bra{\rho_{AB}^{m/2}}\Bigg)^n.
\end{align}
For integer $n$ and even integer $m$, we can compute $Z_{m,n}$ using a path integral on $mn$ copies of the system that are glued together as shown in Figure~\ref{fig:replica}.

We label the individual copies in terms of $(\mu,\nu)\in \mathbb{Z}_m\times \mathbb{Z}_n$.
For each copy, we introduce cuts $A(\mu,\nu)^{\pm}$ and $B(\mu,\nu)^{\pm}$, where $\pm$ indicates the bra/ket portion of the reduced density matrix $\rho_{AB}$.
$A(\mu,\nu)^{+}$ is then identified with $A(\mu,\nu)_{g_A}^{-}$ and $B(\mu,\nu)^{+}$ is identified with $B(\mu,\nu)_{g_B}^{-}$, where $g_A$ and $g_B$ refer to specific permutations that are illustrated in Figure~\ref{fig:replica}.
To be precise, their actions are
\begin{align}\label{eq:perm}
     g_A &: (\mu,\nu)\rightarrow(\mu+1,\nu)\\
      g_B &: (\mu,\nu)\rightarrow(\mu+1,\nu+\delta_{\mu,m/2-1}-\delta_{\mu,m-1})
\end{align}

The partition function $Z_{m,n}$ on the replicated manifold $\mathcal{M}_{m,n}$ can then be analytically continued to obtain the reflected entropy as
\begin{equation}\label{eq:SRmn}
S_R(A:B)=-\underset{m\rightarrow 1}{\text{Lim}}~\partial_n\pa{\frac{\log~Z_{m,n}}{n}}\Big{|}_{n=1}.\end{equation}

\begin{figure}[h]

  \includegraphics[clip,width=0.8\columnwidth]{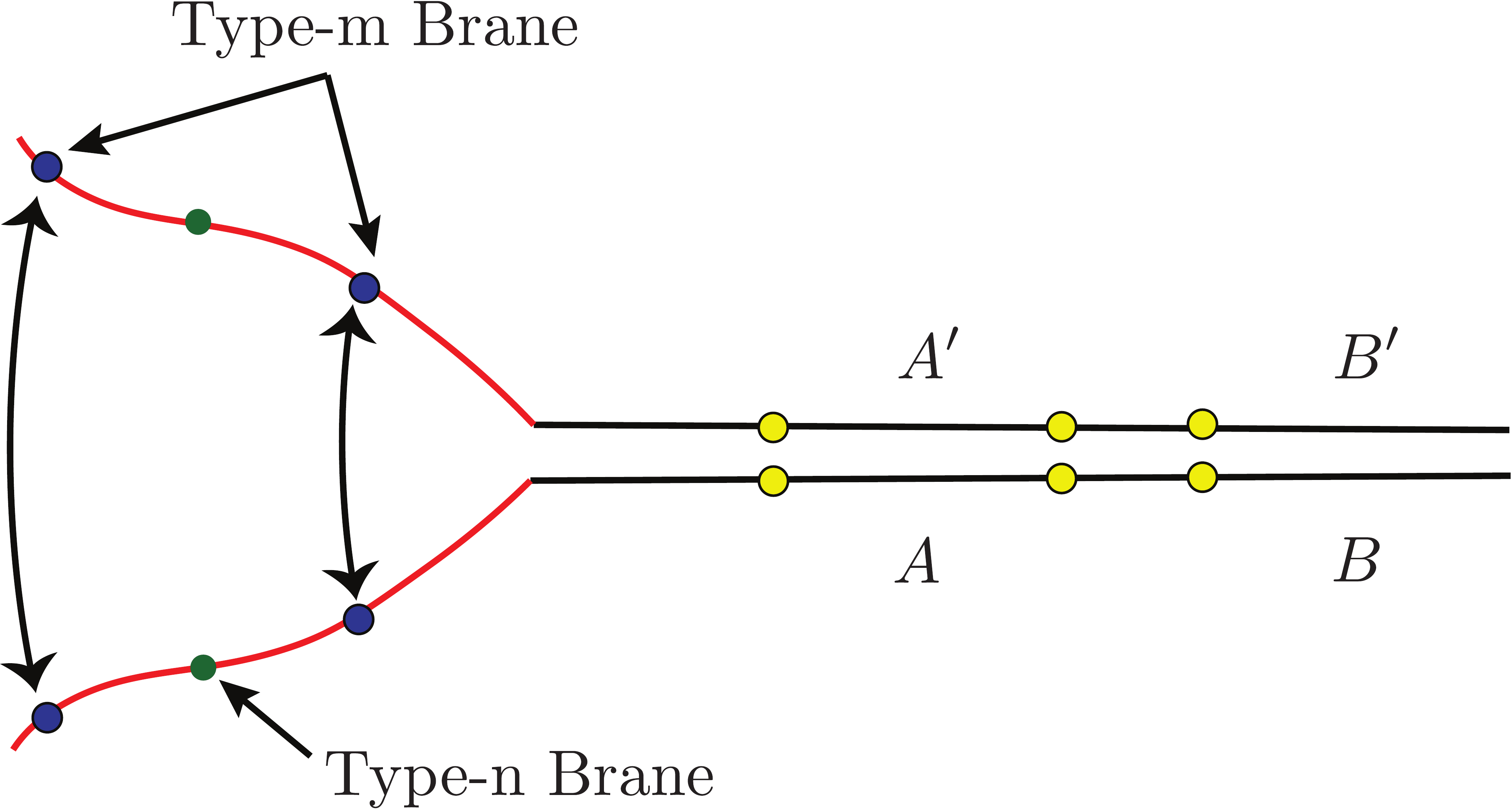}

\caption{The time slice $\Sigma_m$ consists of a gravitating region (denoted red) where two copies of the island region $\text{Is}(AB)$ are glued together at $\partial\text{Is}(AB)$ (denoted purple). The non-gravitating region involves twist operators inserted at $\partial A$ and $\partial B$ (denoted yellow). The effect of these twist operators can be thought of as inducing two kinds of cosmic branes in the gravitating region, which we call Type-$m$ and Type-$n$ branes.}
\label{fig:sigma}
\end{figure}

A convenient way to compute $Z_{mn}$ in a non-gravitational CFT is to use a permutation orbifold theory of $mn$ copies of the CFT, i.e., $\text{CFT}^{\otimes mn}/S_{mn}$, where $S_{mn}$ is the permutation group on $mn$ elements.
With this setup, the partition function $Z_{mn}$ can be computed as
\begin{align}\label{eq:twist}
    Z_{mn} &= \langle \Sigma_{g_A} \Sigma_{g_B}\rangle_{\text{CFT}^{\otimes mn}/S_{mn}}
\end{align}
where $\Sigma_{g_A/g_B}$ represent twist operators that implement the gluing indicated in Figure~\ref{fig:replica}.
$\Sigma_{g_A}$ and $\Sigma_{g_B}$ are operators with conformal dimensions proportional to $m-1$, we shall refer to these as Type-$m$ twist operators.

In this situation it is harder to perform an analytic continuation directly in the bulk, but it is useful to have a heuristic picture of the relevant physics.
When coupled to gravitational regions, the Type-$m$ twist operators spontaneously create conical defects with opening angle $\frac{2\pi}{m}$ that we term Type-$m$ branes as seen in Figure~\ref{fig:sigma}.
These Type-$m$ branes demarcate the island region $\text{Is}(AB)$ associated as noted in \cite{Almheiri:2019qdq}.
Another useful operator to consider is $\Sigma_{g_A\,g_B^{-1}}$, the dominant operator exchanged between $\Sigma_{g_A}$ and $\Sigma_{g_B^{-1}}$.
It has conformal dimension proportional to $n-1$ and we shall refer to it as a Type-$n$ twist operator.
These induce Type-$n$ branes with opening angle $\frac{2\pi}{n}$, which land on the cross section of the island, and their contribution leads to our formula in Eqn.~(\ref{eq:islandSR}).
This discussion makes it clear that in order to avoid the branes backreacting on each other, we must consider the limit where $n\rightarrow1$ first so that the Type-$n$ branes can be treated as probes.

\section{Phase transitions}\label{sec:examples}

In this section, we consider phase transitions of the reflected entropy in JT gravity coupled to a bath. 
In the gravitating $\text{AdS}_2$ region we have a $\text{CFT}_2$ eternally coupled to a $\text{CFT}_2$ in the Minkowski region, the latter of which serves as a ``bath system''. 
The full action of the theory is 
\begin{align}
    I = \frac{1}{4\pi}\int d^2 x \sqrt{-g}[\phi R + 2(\phi - \phi_0)] + I_{\text{CFT}}
\end{align}
where we follow the conventions of \cite{Almheiri:2019yqk} in which $4G_N = 1$. Here $\phi_0$ is the extremal value of the dilaton $\phi$.  
We will consider two classes of examples, where the gravity theory is in equilibrium with the bath system, at either zero or finite temperature.

\subsection*{Vacuum $\text{AdS}_2$}

\begin{figure}[h]

  \includegraphics[clip,width=0.8\columnwidth]{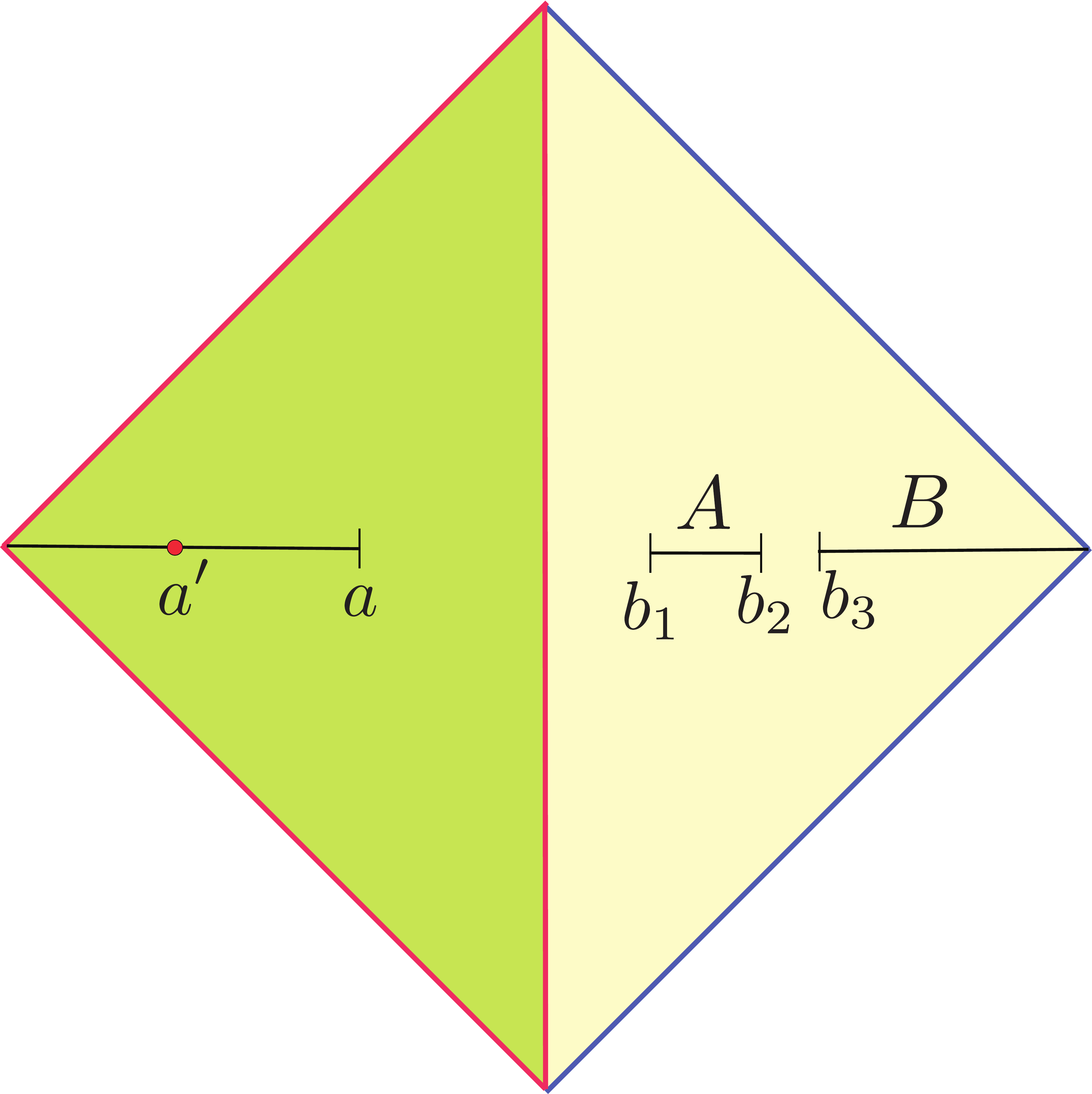}

\caption{The Penrose diagram for the vacuum $\text{AdS}_2$ setup consisting of a finite subregion $A$ and a semi-infinite subregion $B$ in a half-Minkowski space (bath) eternally coupled to a gravitating region with the correspond island $a$ and cross-section $a'$.}
\label{fig:zerotemp}
\end{figure}

Here, we consider a vacuum $\text{AdS}_2$ solution glued to a half Minkowski space as seen in Figure~\ref{fig:zerotemp}.
This can be thought of as the zero temperature limit of an eternal black hole.
The metric and dilaton profile in the $\text{AdS}_2$ region are 
\begin{align}
    ds^2 &= -4\frac{dx^+ dx^-}{(x^- - x^+)^2} \label{metric} \\
    \phi(x) &= \phi_0 + 2\frac{\phi_r}{x^- - x^+} \\ 
    x^{\pm} &= t \pm z, \ z\in (-\infty, 0] 
\end{align}
In the flat space region the metric is the standard Minkowski metric, 
\begin{align}
    ds^2 = -dt^2 + dz^2 , \ z\in [0,\infty) 
\end{align}
Given this setup, we will consider two different choices of subregions that will give us qualitatively different behaviour.

\subsubsection*{Example 1}

The first example we consider involves two intervals in the bath region, $[b_1,b_2]$ and $[b_3,\infty)$ where $b_3 > b_2$. 
Let $(-\infty,-a]$ denote the corresponding island in the $\text{AdS}_2$ region.

We now compute the reflected entropy $S_R(A:B)$ for this setup. The calculation will be similar to that of \cite{Almheiri:2019yqk}, while making use of the techniques in \cite{Dutta:2019gen}. To start with, the effective reflected entropy, is obtained from the $n\rightarrow 1$ limit of  \cite{Dutta:2019gen} 
\begin{align}
    S_n(\rho^{m/2}_{AB}) = \frac{1}{1-n} \log \frac{\langle \Sigma_{g_A}\Sigma_{g_B}\rangle}{\langle \Sigma_{g_m}\Sigma_{g_m}\rangle^n} \label{effref}
\end{align}
where the operator dimensions are 
\begin{align}
    \Delta_{g_A} = \Delta_{g_B} = \frac{cn(m^2-1)}{24m} = n\Delta_m 
\end{align}
We now point out a useful property of the reflected entropy under Weyl transformations. Consider a Weyl transformation $g \rightarrow \Omega^2 g$ of the metric. Then, 
\begin{align}
    \langle \Sigma_{g_A}\Sigma_{g_B}\rangle &\rightarrow \Omega_A^{\Delta_A}\Omega_B^{\Delta_B}\langle \Sigma_{g_A}\Sigma_{g_B}\rangle \\
    \langle \Sigma_{g_m}\Sigma_{g_m}\rangle^n &\rightarrow (\Omega_A \Omega_B)^{n\Delta_m}\langle \Sigma_{g_m}\Sigma_{g_m}\rangle^n
\end{align}
Hence the Weyl factors cancel out in Eqn.~ \eqref{effref} for coincident points in the numerator and denominator. In general, if there is no island then the reflected entropy will be entirely Weyl invariant. If there is an island with a cross-section then the twist insertion at the cross-section will have an additional Weyl factor. This simplifies the analysis, since we can simply do the entire calculation in flat space by an appropriate Weyl transformation of Eqn.~ \eqref{metric}. 

\begin{figure}[h]

  \includegraphics[clip,width=0.8\columnwidth]{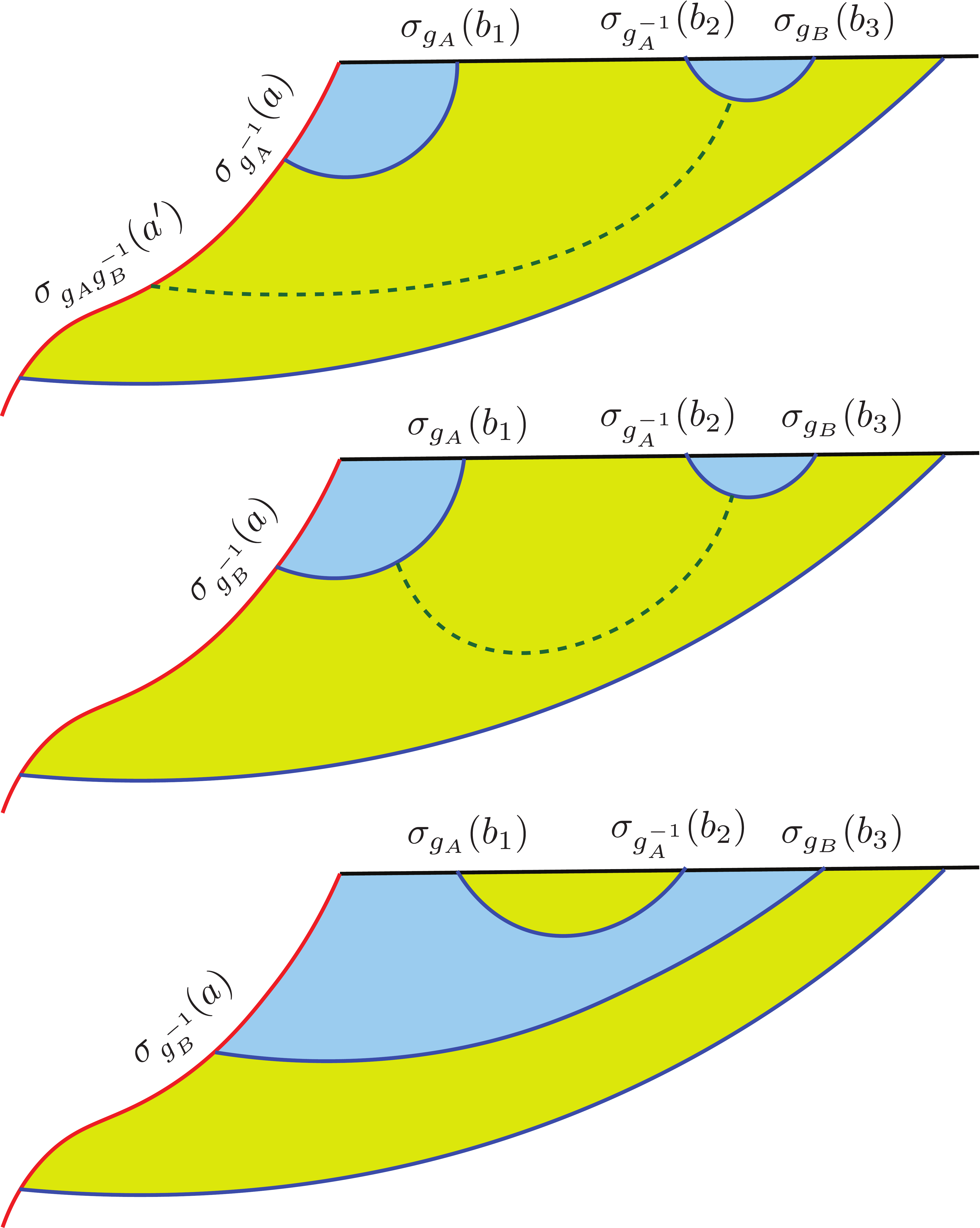}

\caption{The three possible phases along with the associated contractions of twist operators. \emph{Top}: connected phase of the entanglement island, with a non-trivial cross-section. \emph{Middle}: connected phase of the entanglement island, with no cross-section. \emph{Bottom:} disconnected phase of the entanglement island.}
\label{fig:OPE}
\end{figure}

Let us first review the salient aspects of computing $S_n(\rho^{m/2}_{AB})$ for a $\text{CFT}_2$ in flat space as described in \cite{Dutta:2019gen}. In this case the twist operator $\Sigma_g$ associated to an interval become two quasi-local twist operators $\sigma_g, \sigma_{g^{-1}}$ inserted at the endpoints of the interval. The dominant exchange between $\sigma_{g_A}$ and $\sigma_{g_B^{-1}}$ is a composite operator denoted $\sigma_{g_Ag_B^{-1}}$ which has dimension 
\begin{align}
\Delta_{g_Ag_B^{-1}} = \frac{c}{12n}(n^2-1) = 2\Delta_n 
\end{align}
The OPE coefficient of the exchange is 
\begin{align}
    C_{n,m} = (2m)^{-4\Delta_n } 
\end{align}

We can now do the calculation for the present setup. Let $a'$ denote the cross-section of the island. We must evaluate 
\begin{align}
    S_{n,m} = \frac{1}{1-n}\log \frac{\langle \sigma_{g_A}(b_1)\sigma_{g_A^{-1}}(a)\sigma_{g_A^{-1}}(b_2)\sigma_{g_B}(b_3)\sigma_{g_{A}g_B^{-1}}(a') \rangle_{mn}}{\langle\sigma_{g_m}(b_1)\sigma_{g_m^{-1}}(a)\sigma_{g_m^{-1}}(b_2)\sigma_{g_m}(b_3) \rangle_m^n} \label{effrelcalc}
\end{align}
Note that once we map the $\text{AdS}_2$ region to flat space, all Weyl factors will cancel out except the one associated with the cross-section $a'$. 

The contractions corresponding to the first phase are shown in Figure \ref{fig:OPE}. At this point we restrict to a large $c$ CFT, in which case the correlators factorize into their respective contractions \cite{hartman2013entanglement}, hence we get \begin{align}
    &\langle \sigma_{g_A}(b_1)\sigma_{g_A^{-1}}(a)\sigma_{g_A^{-1}}(b_2)\sigma_{g_B}(b_3)\sigma_{g_{A}g_B^{-1}}(a') \rangle_{mn} \\
    &= \langle\sigma_{g_A}(b_1)\sigma_{g_A^{-1}}(a)\rangle_{mn}\langle \sigma_{g_A^{-1}}(b_2)\sigma_{g_B}(b_3)\sigma_{g_{A}g_B^{-1}}(a')\rangle_{mn}
    \\ &= \Omega^{2\Delta_n}(a')\frac{1}{(a + b_1)^{2n\Delta_m}} \frac{C_{n,m}}{(b_3 - b_2)^{2n\Delta_m - 2\Delta_n} (b_3+a')^{2\Delta_n}}\frac{1}{(b_2 + a')^{2\Delta_n}}
\end{align}
Similarly, 
\begin{align}
\langle\sigma_{g_m}(b_1)\sigma_{g_m^{-1}}(a)\sigma_{g_m^{-1}}(b_2)\sigma_{g_m}(b_3) \rangle_m = \frac{1}{(a+b_1)^{2\Delta_m}}\frac{1}{(b_3-b_2)^{2\Delta_m}}    
\end{align}
Putting this together, the reflected entropy of this phase is 
\begin{multline}
S^{(1)}_R = \frac{c}{3}\left(\log(b_3 + a')+ \log(b_2 + a') -\log 4a' -\log(b_3-b_2)\right)\\
+ 2\frac{\phi_r}{a'}+2\phi_0 \label{refent}
\end{multline}
The last term will be divergent in the limit $b_2 \rightarrow b_3$. However, we are only interested in extracting the phase transition in $S_R$. Therefore this divergence will not matter; as we will see below, the same divergent term appears in the reflected entropy of the other phase. 

The location of $a'$ is then determined from the extremization 
\begin{align}
    \frac{dS^{(1)}_R}{da'} = 0 
\end{align}
which yields the cubic equation
\begin{align}
    \frac{c}{3}\left(\frac{1}{b_3 + a'} + \frac{1}{b_2 + a'} - \frac{1}{a'} \right) - 2\frac{\phi_r}{(a')^2} = 0 \label{solvevac}
\end{align}
This is a cubic equation for $a'$ which can be solved in principle but is not quite illuminating. Instead we will consider the simplifying limit $b_2 \rightarrow b_3$. Then, Eqn.~\eqref{solvevac} simplifies to the quadratic
\begin{align}
    (a')^2 - a'(b_3 + 6\frac{\phi_r}{c}) - 6\frac{\phi_r}{c}b_3 = 0
\end{align}
whose solution is 
\begin{align}
    2a' = b_3 + 6\frac{\phi_r}{c} + \sqrt{b_3^2 + 36\frac{\phi_r}{c}b_3 + 36\frac{\phi_r^2}{c^2}}
\end{align}

In order to analyze the phase transition, we also need the reflected entropy for the second non-trivial phase (see Figure \ref{fig:OPE}). This is the phase where there is no cross-section $a'$ so we just compute 
\begin{align}
    S^{(2)}_R = \frac{1}{1-n}\log \frac{\langle \sigma_{g_B^{-1}}(a)\sigma_{g_A}(b_1)\sigma_{g_A^{-1}}(b_2)\sigma_{g_B}(b_3)\rangle_{mn}}{\langle\sigma_{g_m^{-1}}(a)\sigma_{g_m}(b_1)\sigma_{g_m^{-1}}(b_2)\sigma_{g_m}(b_3) \rangle^n_m }
\end{align}
This calculation is identical to the one done in \cite{Dutta:2019gen}, and the result is 
\begin{align}
    S^{(2)}_R = \frac{2c}{3}\log \left(1 + \frac{\sqrt{1-x}}{\sqrt{x}} \right), \ x = \frac{(b_3-b_2)(b_1 + a)}{(b_3 - b_1)(b_2 + a)}
\end{align}
In the limit $b_2 \rightarrow b_3$, the cross ratio $x \rightarrow 0$. Thus, 
\begin{align}
    S_R^{(2)} \approx \frac{c}{3}\left(\log(b_3 + a) + \log (b_3 - b_1) -\log(b_1 +a) - \log(b_3 - b_2) \right) \label{phasetwo}
\end{align}
As mentioned, we see that the same divergent term appears in both $S_R^{(1)}$ and $S_R^{(2)}$. 

The expression for $a$ was obtained from the standard islands prescription in \cite{Almheiri:2019yqk} and the result is  
\begin{align}
    2a(b_1) = b_1 + 6\frac{\phi_r}{c}+\sqrt{b_1^2 + 36 \frac{\phi_r}{c}b_1 + 36\frac{\phi_r^2}{c^2}} \label{islandval}
\end{align}
which is valid for the connected phase of the entanglement island.
Note that we have made the dependence $a(b_1)$ explicit since similar expressions will appear in other calculations too.

Lastly, we have the third phase in Figure \ref{fig:OPE}. This corresponds to the disconnected phase of the entanglement island, for which 
\begin{align}
S^{(3)}_R = 0
\end{align}
For this phase, we simply have $a(b_3)$ instead of $a(b_1)$ in Eqn.~\eqref{islandval}. 

To summarize, the first two phases correspond to the case where the entanglement island of $AB$ is connected, whereas in the third phase the entanglement island of $AB$ is disconnected. The phase transition can then be analyzed in the following series of steps: 
\begin{itemize}
    \item Find the phase transition in the entanglement entropy, which corresponds to a transition in the entanglement island of $AB$ 
    \item Within each phase of the entanglement entropy, find the phase transition in the reflected entropy 
\end{itemize}

To do this analysis analytically, we first work in the limit where $\phi_r/c \gg 1$. At leading order, for \emph{both} phases, we then have  
\begin{align}
    a \approx  \frac{6\phi_r}{c} 
\end{align}

Then at leading order the phase transition in the entanglement entropy occurs when \begin{align}
    b_2 \approx \frac{b_1 + b_3}{2}
\end{align}
When $b_2$ is below this value we are in the disconnected phase, and when it is above this value we are in the connected phase. Note that in the limit $b_2 \rightarrow b_3$ we will always be in the connected phase. 

In the simultaneous limits of large $\phi_r/c$ and $b_2 \rightarrow b_3$, we have 
\begin{align}
    a' \approx 6\frac{\phi_r}{c}
\end{align}

We then see from Eqns.~\eqref{refent} and \eqref{phasetwo} that the following behavior holds: 
\begin{align}
    S_R^{(1)} &\sim \frac{c}{3}\log \frac{\phi_r}{c} + 2\phi_0 \\ 
    S_R^{(2)} &\sim \frac{c}{3}\log (b_3-b_1) 
\end{align}
Since $\phi_r/c \gg b_3,b_1$ is the largest scale in our parameter regime, we see that $S_R^{(1)} > S_R^{(2)}$ always. Thus, there is no phase transition within the connected entanglement island phase. Instead, we are always in the second phase in Figure \ref{fig:OPE}. Physically, the subsystem $A$ is too small to accommodate large bi-partite quantum entanglement with $B$, so that the entanglement island does not contribute to bi-partite correlation measures.

We now consider the opposite limit, $b_2 \rightarrow b_3 \gg \phi_r/c$. In this parameter regime, we have  
\begin{align}
    a' &\approx b_3 \\
    a &\approx b_1
\end{align}
Consequently, the reflected entropies go like 
\begin{align}
    S_R^{(1)} &\sim \frac{c}{3}\log b_3 + 2\phi_0 \\
    S_R^{(2)} &\sim \frac{c}{3}\log b_3^2 
\end{align}
Thus we see that there is a phase transition when 
\begin{align}
    b_3 \sim e^{6\phi_0/c}
\end{align}
This example demonstrates the fact that even the entanglement between subsystems in the Hawking radiation gets large modifications from gravity, provided that there is large entanglement.

In the next example, we consider a setup where one subsystem contains the black hole and the other is a subsystem in the bath.

\subsubsection*{Example 2}

\begin{figure}[h]

  \includegraphics[clip,width=0.8\columnwidth]{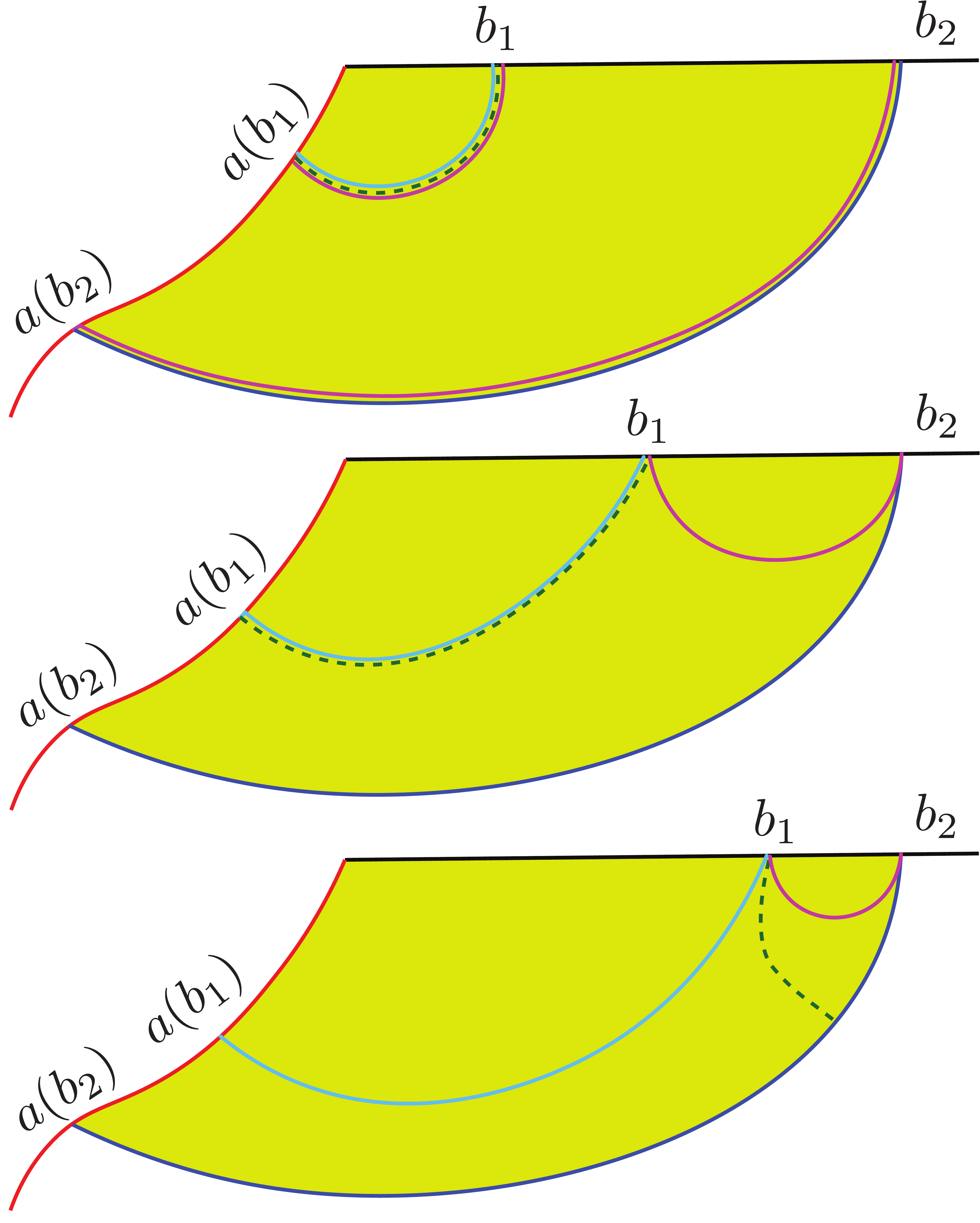}

\caption{As we vary $b_1$ we see three possible phases based on the behaviour of the various surfaces in the double holography picture, RT surface of $A$ (light blue), RT surface of $B$ (pink) and the entanglement wedge cross section (green dashed line).}
\label{fig:transition}
\end{figure}

\begin{figure}[h]

  \includegraphics[clip,width=\columnwidth]{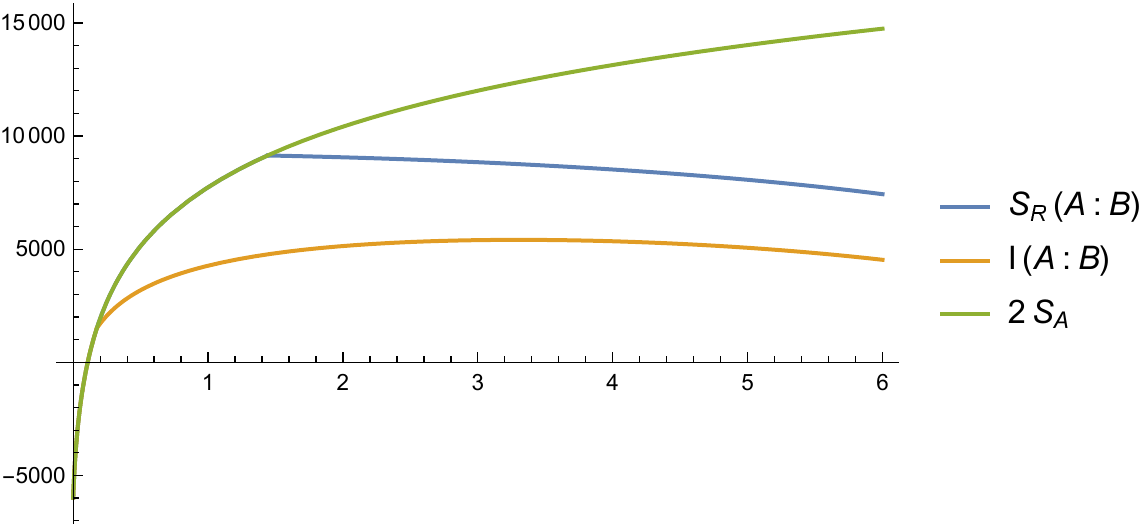}

\caption{We plot the behaviour of $2S(A)$, $S_R(A:B)$ and $I(A:B)$ as a function of $b_1$, for $b_2=10$, $\phi_0=1000$. $\phi_r=100$, and $c=12000$. The phase transition of $S_R$ is in accord with eq.(\ref{eq:transition}).}
\label{fig:plot}
\end{figure}

The second example we consider is of subregions that are adjacent intervals, $A:=[0,~b_1]$ and $B:=[b_1,~b_2]$. $A$ now contains QM system which is dual to JT gravity.
We compute the reflected entropy as a function of $b_1$ holding $b_2$ fixed.
We first describe the phase transition qualitatively as can be well understood in terms of the double holography picture seen in Figure~\ref{fig:transition}.
We then plot the behaviour quantitatively in Figure~\ref{fig:plot}.

As found in \cite{Almheiri:2019yqk}, the subregion $A$ always includes an entanglement island and its von Neumann entropy is given by
\begin{equation}
S(A)=S_{\text{gen}}(b_1):=\phi_0+\frac{\phi_r}{a}+\frac{c}{6}\log\frac{(a(b_1)+b_1)^2}{a(b_1)\epsilon},
\end{equation}
where $[-a(b_1),0]$ is the corresponding island where $a(b_1)$ is given by Eqn.~\eqref{islandval}.
In various limits, it is approximately given by
\begin{align}
    a(b_1)\approx \begin{cases} \frac{6\phi_r}{c} &\mbox{if } b_1 \ll \frac{\phi_r}{c} \\ 
b_1 & \mbox{if } b_1 \gg \frac{\phi_r}{c} \end{cases}.
\end{align}
For the subregion $B$, it was shown in \cite{Almheiri:2019yqk} that it contains an island for small $b_1$ only if $b_2 > \frac{\phi_r}{c} \exp(12\phi_0/c)$.
We consider fixed but large $b_2$ such that $B$ can have an entanglement island. 

For small $b_1$, it is preferable for $B$ to have an entanglement island leading to
\begin{align}
S(B)&=S_{\text{gen}}(b_1)+S_{\text{gen}}(b_2)\\
I(A:B)&=2S_{\text{gen}}(b_1)=2S(A).
\end{align}
For large $b_1$, $B$ does not have an entanglement island, and we instead have
\begin{align}
    S(B)&=\frac{c}{3}\log\frac{b_2-b_1}{\epsilon}.
\end{align}
Thus, in this limit we obtain
\begin{equation}
I(A:B)=S_{\text{gen}}(b_1)+\frac{c}{3}\text{log}\frac{b_2-b_1}{\epsilon}-S_{\text{gen}}(b_2).
\end{equation}

In order to compute the reflected entropy, we can now include various reflected entropy islands as dictated by our proposed formula, Eqn.~(\ref{eq:islandSR2}).
For small $b_1$, there is a non-trivial splitting of the entanglement island as seen in Figure~\ref{fig:transition}.
This calculation can be done using the techniques employed in the previous section.
In the end, the extremization to obtain the reflected entropy island ends up being identical to that in the entanglement island calculation.
Hence, we obtain
\begin{equation}
S_R(A:B)=2S_{\text{gen}}(b_1)=2S(A).
\end{equation}
At large $b_1$, there is no non-trivial reflected entropy island i.e. the entire island belongs to $A$.
Again using techniques similar to the ones used in the previous example, we obtain
\begin{equation}
S_R(A:B)=\frac{c}{3}\log\frac{2(a(b_2)+b_1)(b_2-b_1)}{(a(b_2)+b_2)\epsilon},
\end{equation}
where $a(b_2)$ is the location of the island for $B$.
The phase transition between these two phases of the reflected entropy occurs at 
\begin{equation}
\text{e}^{\frac{6\phi_0}{c}}\approx \frac{(b_2+b_1)(b_2-b_1)}{4b_1b_2}.\label{eq:transition}
\end{equation}
We plot this behaviour quantitatively in Figure~\ref{fig:plot}. 

\subsection*{Eternal Black Hole}

\begin{figure}[h]

  \includegraphics[clip,width=\columnwidth]{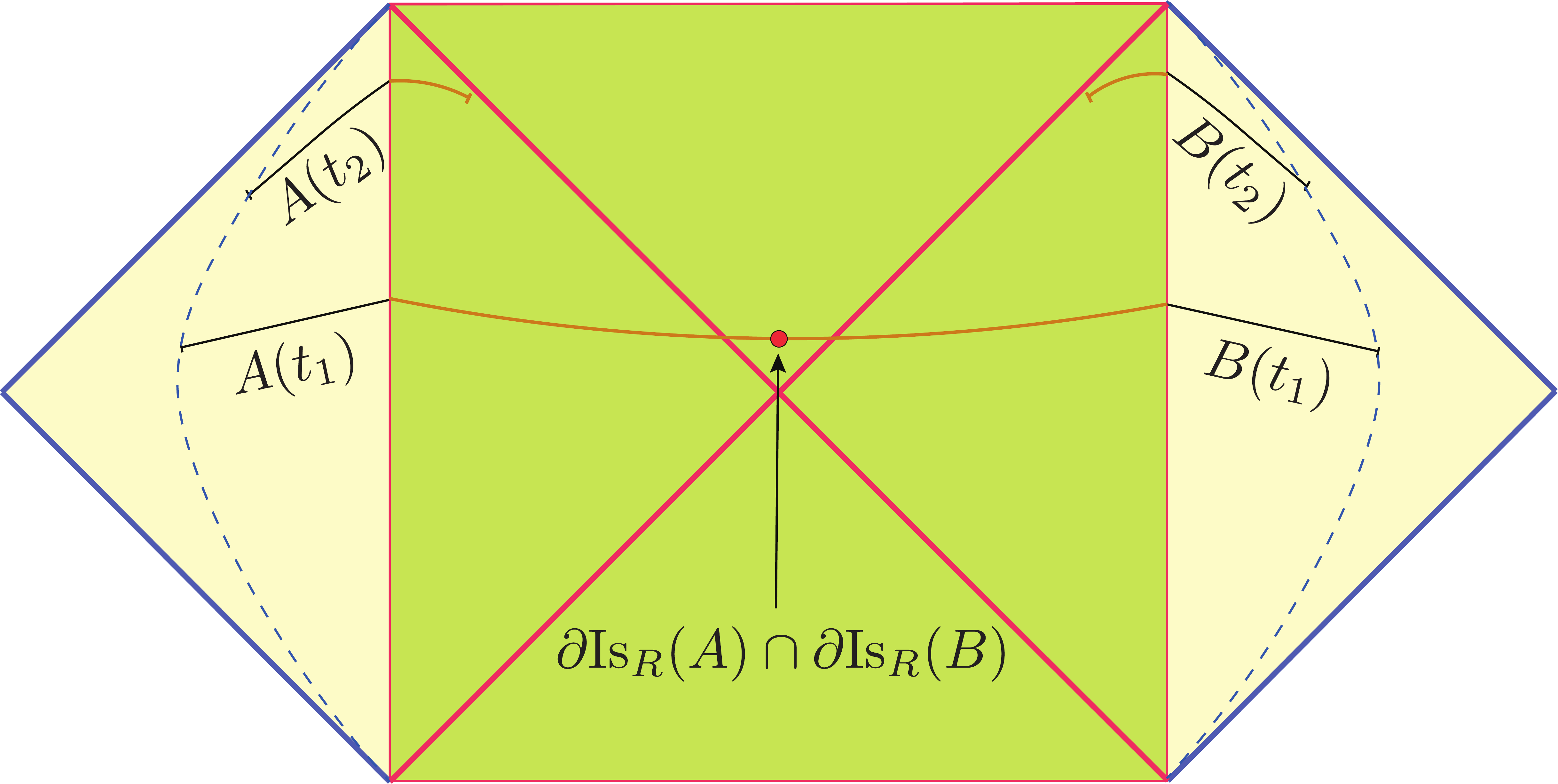}

\caption{The eternal black hole coupled to a bath CFT in Minkowski space is considered with subregions $A$ and $B$ at different times. At early times, the subregion $AB$ has an entire Cauchy slice of the gravity region as its entanglement island (denoted orange). The computation of $S_R(A:B)$ then includes an area contribution from the boundary of the reflected entropy island (denoted red). At late times, the entanglement island is disconnected and $S_R(A:B)=0$.}
\label{fig:finitetemp}
\end{figure}

Having considered the vacuum $\text{AdS}_2$ example, we now come to the example of a two-sided eternal black hole glued to two halves of Minkowski space, one on each side. The full $\text{CFT}_2$ is in the Hartle-Hawking state, as considered in \cite{Almheiri:2019yqk}. See Figure \ref{fig:finitetemp} for the Penrose diagram of the setup. The metric and dilaton profile for each black hole exterior and respective bath are 
\begin{align}
    ds^2 &= -\frac{4\pi^2}{\beta^2}\frac{dy^+ dy^-}{\sinh^2\frac{\pi}{\beta}(y^- - y^+)} \\
    \phi&= \phi_0 + \frac{2\pi \phi_r}{\beta}\frac{1}{\tanh\frac{\pi}{\beta}(y^- - y^+)}
\end{align}
where the $(y^+, y^-)$ coordinates are related to the $(x^+,x^-)$ coordinates via 
\begin{align}
    x^{\pm} = \tanh \frac{\pi y^{\pm}}{\beta}
\end{align}

The calculation will proceed similarly to that of the previous section. Namely, we do a Weyl transformation to map the black hole geometry to flat space and apply standard $\text{CFT}_2$ techniques to calculate the reflected entropy in the presence of an island. As before, we work in the large $c$ limit throughout, so that we can factorize correlators into their contractions. 

At $t=0$ we consider an interval $[-b,0]$ in the left Minkowski region and an interval $[0,b]$ in the right Minkowski region. In this symmetric setup, the corresponding ``island" in the black hole spacetime will be the entire bulk slice, with a cross-section $a'$ splitting it, see Figure \ref{fig:finitetemp}. This is the early time phase. Just as in \cite{Almheiri:2019yqk} we then move both endpoints of the bath intervals forward in time, which introduces time-dependence into the setup. The resulting reflected entropy will therefore be a function of time. At late times there will be phase transition to a disconnected phase depicted in Figure \ref{fig:finitetemp}. 

The rule as prescribed in our formula, Eqn.~\ref{eq:islandSR2}, is to first find the phase transition in the entanglement entropy.
Given the entanglement island, we can then find the phase transitions in the reflected entropy for subregions which all share the same island. 
In the present case the phase transition occurs entirely at the level of the entanglement entropy. In other words, the phase transition in $S_R(A:B)$ is completely dictated by the entanglement entropy phase transition. 

The entanglement entropy $S(AB)$ grows linearly with time similar to the calculation in \cite{Hartman:2013qma}, and eventually has a phase transition as demonstrated in \cite{Almheiri:2019yqk}.
Note that our setup is simply the complement of the setup in \cite{Almheiri:2019yqk}. 
Hence the entanglement entropy phase transition happens at the same time, given by
\begin{align}
t \sim \frac{\beta S_{\text{BH}}}{c}
\end{align} 
where $S_{\text{BH}} = 2(\phi_0 + 2\pi \phi_r/\beta)$ is the black hole entropy. This is therefore also the time at which the reflected entropy phase transition happens. 

In order to compute the reflected entropy for the early time phase, we need to convert to global coordinates that cover the entire spacetime. Let $y^{\pm}_L$ denote the coordinates covering the left exterior and bath, and similarly for $y^{\pm}_R$. Then  
\begin{align}
    w^{\pm} = \pm e^{\pm \frac{2\pi y^{\pm}_R}{\beta}}, \ w^{\pm} = \mp e^{\mp \frac{2\pi y^{\pm}_L}{\beta}}
\end{align}
defines a global coordinate chart $(w^+, w^-)$ in which the metric simply reduces to that of Minkowski space. Let $w^{\pm}_1$ denote $a'$, $w^{\pm}_2$ denote the endpoint of the left bath interval, and $w^{\pm}_3$ denote the endpoint of the right bath interval.
Moreover, recall the relations 
\begin{align}
    y_L^{\pm} = t\mp z, \ y_R^{\pm} = t\pm z 
\end{align}
where $z$ denotes the spatial coordinate. Lastly, by symmetry we have that $w_1^+ = w_1^- := \delta$. We can now compute the \emph{effective} reflected entropy for this phase, which is given by 
\begin{align}
    S_R^{\text{(eff)}} &= -\lim_{n\rightarrow 1} \partial_n\left[\frac{c}{12}\frac{(n-\frac{1}{n})}{n}\log \frac{\varepsilon^2 (1 + w_1^+w_1^-)^2}{4\left(\frac{w_{12}^+w_{12}^-w_{13}^+w_{13}^-}{w_{23}^+w_{23}^-} \right)} + \frac{1}{n}\log C_n^2  \right]
    \\ &= \frac{c}{3}\log 2 + \frac{c}{6}\log \varepsilon^2\frac{e^{4\pi b/\beta}(e^{2\pi t/\beta} - \delta e^{-2\pi b/\beta})^2(\delta e^{-2\pi b/\beta} +e^{-2\pi t/\beta})^2}{(1+\delta^2)\cosh^2 2\pi t/\beta}
\end{align}
where $\varepsilon$ is the UV cutoff. The reflected entropy of this phase is therefore 
\begin{align}
    S^{(1)}_R = \text{ext}_\delta \left(S_R^{\text{(eff)}} + 2\left(\phi_0 + \frac{2\pi \phi_r}{\beta}\frac{1-\delta^2}{1+\delta^2}\right) \right)
\end{align}
In the limit $b\rightarrow \infty$, the solution to 
\begin{align}
    \frac{dS^{(1)}_R}{d\delta} = 0
\end{align}
is simply $\delta \rightarrow 0$ which is a consequence of the symmetry in the setup. 

Thus, the final expression for the reflected entropy of the non-trivial phase in this limit is 
\begin{align}
    S^{(1)}_R = 2(\phi_0 + \frac{2\pi \phi_r}{\beta})+ \frac{c}{3}\log 2 -\frac{c}{6}\log \frac{\cosh^2 2\pi t/\beta}{\varepsilon^2 e^{4\pi b/\beta}}
\end{align}
Thus, we see that at early times the reflected entropy decreases linearly just like the mutual information \cite{Almheiri:2019qdq}.
On the other hand, the reflected entropy for the late time phase clearly vanishes, 
\begin{align}
S^{(2)}_R = 0.
\end{align}
Since the transition is completely dictated by the entanglement entropy phase transition, generically there is a discontinuous jump in $S_R(A:B)$ from a non-zero value to zero when the transition occurs.
This is analogous to the situation in the phase transition for two intervals in pure AdS \cite{Umemoto:2019jlz,Akers:2019gcv}.


\section{Discussion}\label{sec:disc}
\subsection*{Generalizations}
In this work, we have proposed a formula in Eqn.~(\ref{eq:islandSR}) that generalizes the holographic conjecture for reflected entropy to situations where the entanglement wedge includes an island.
There have also been holographic proposals for other measures of bipartite correlation such as entanglement negativity, odd entropy and entanglement wedge mutual information\cite{Chaturvedi:2016rcn,Rangamani:2014ywa,Tamaoka:2018ned,Kudler-Flam:2018qjo,Kusuki:2019zsp, Umemoto:2019jlz}.
Similarly, multipartite versions of the reflected entropy and entanglement of purification have also been proposed previously \cite{Bao:2018gck,Bao:2019zqc,Marolf:2019zoo}.
Although we have focused on the reflected entropy in this paper, the generalization essentially can be understood as applying the usual holographic formula after including the island in the entanglement wedge. Hence, all the proposals discussed above can also be generalized in a similar manner to include island contributions. We expect such generalizations will help understanding the entanglement structure of Hawking radiation, while it is also important to understand physical implications of those correlation measures. 

\subsection*{Interpretation of Results}

We now attempt to interpret some of the results we have obtained in Section~\ref{sec:examples}.
The reflected entropy is known to be a correlation measure that includes contributions from both classical and quantum correlations \cite{Dutta:2019gen,Umemoto:2019jlz,Levin:2019krg,Kudler-Flam:2020url}.
However, there isn't a general understanding of the precise kind of correlations quantified by the reflected entropy, and also no general classification of multipartite entanglement.
Hence, we will use suggestive examples that exhibit features similar to those found in Section~\ref{sec:examples}.

The interesting features of the examples in Section~\ref{sec:examples}, especially the behaviour found in Figure~\ref{fig:plot}, can be understood from the interplay between the reflected entropy $S_R(A:B)$, and its upper and lower bounds given by
\begin{align}\label{eq:bounds}
    \text{min}\{2 S(A),2 S(B)\} & \geq S_R(A:B) \geq I(A:B).
\end{align}
In certain regions of parameter space, we see saturation of the upper and lower bounds and we can take this as a guiding principle to indicate the structure of entanglement of the states.
We will demonstrate these features in two examples and conjecture that this is indicative of the general correlations quantified by $S_R(A:B)$.

\subsubsection*{Qubit example}

The first example comprises a three party qubit state, where the multipartite entanglement structure is quite well understood \cite{Dur:2000zz,Rangamani:2015qwa}.
The simplest form of entanglement is Bell pair like bipartite entanglement between any two of the parties, e.g.,
\begin{align}
    \ket{\text{Bell}}&=\frac{1}{\sqrt{2}}(\ket{00}+\ket{11})\otimes\ket{0}.
\end{align}
Genuine tripartite entangled states can be classified into two categories represented by the GHZ state and W state given by
\begin{align}
    \ket{\text{GHZ}}&=\frac{1}{\sqrt{2}}(\ket{000}+\ket{111})\\
    \ket{\text{W}}&=\frac{1}{\sqrt{3}}(\ket{001}+\ket{010}+\ket{100})
\end{align}
Heuristically the difference between these two kinds of states is that any two parties are only classically correlated in the GHZ state, whereas they have genuine quantum correlation in a W state.

A simple computation tells us that the Bell state saturates both upper and lower bounds in Eqn.~(\ref{eq:bounds}).
Similarly, the GHZ state saturates only the lower bound whereas the W state saturates neither.
This gives us a heuristic picture that non-saturation of the upper bound is related to the existence of tripartite entanglement, whereas non-saturation of the lower bound is related to the existence of quantum correlation \footnote{Note that the Bell state has to be treated carefully in a limiting procedure since it saturates both bounds.}.

\begin{figure}[h]

  \includegraphics[clip,width=\columnwidth]{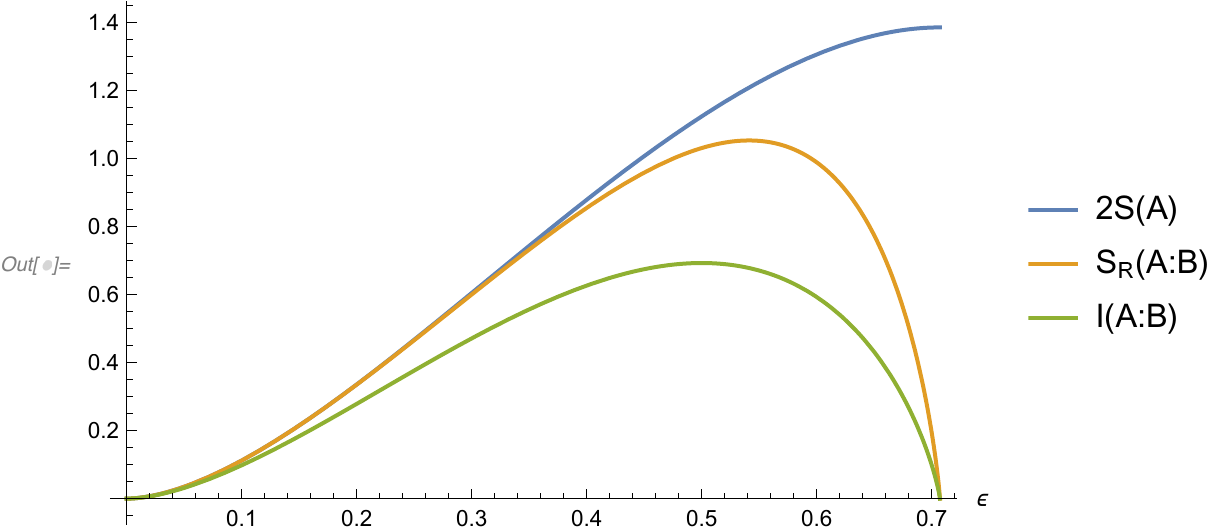}

\caption{The reflected entropy for the state $\ket{W_{\epsilon}}$ as a function of $\epsilon$ is compared to its upper and lower bounds. We see that it has qualitative features similar to that found in Section~\ref{sec:examples}.
}
\label{fig:Weps}
\end{figure}

Given this discussion, we can discuss the structure of entanglement found in Figure~\ref{fig:plot}.
Clearly, the region before the first phase transition corresponds to Bell pair like entanglement as seen from saturation of both bounds.
So we now focus on the region after the first phase transition where the upper bound is saturated but the lower bound is not.
A simple model that captures this feature is a state that can be thought of as a W-like perturbation to a product state which we denote as $\ket{W_{\epsilon}}$,
\begin{align}
    \ket{W_{\epsilon}}&=\epsilon \ket{100} +\epsilon \ket{001} + \sqrt{1-2 \epsilon^2} \ket{010}.
\end{align}
In Figure~\ref{fig:Weps}, we plot the three relevant quantities of this state discussed in Eqn.~(\ref{eq:bounds}) and it can easily be seen that for a significant neighbourhood near $\epsilon=0$, the state saturates the upper bound while staying far away from the lower bound.
More precisely, we find that perturbatively around $\epsilon=0$, we have
\begin{align}
    |S_R(A:B)-I(A:B)|&=O(\epsilon^2)\\
    |S_R(A:B)-2 S(A)|&=O(\epsilon^4 \log(\epsilon))
\end{align}

This hints at the fact that $S_R(A:B)$ is in fact more sensitive to the existence of quantum correlations than it is to the existence of tripartite entanglement of the $\ket{W}$ type.
The importance of W-type tripartite entanglement for holographic states was emphasized in \cite{Akers:2019gcv}.
Going beyond qubit systems, it is unclear what exactly we mean by W-type entanglement since the multipartite entanglement classification becomes much more complicated \cite{verstraete2002four, Rangamani:2015qwa}.
In the context of reflected entropy, what we really mean is states that are far from saturating the lower bound in Eqn.~(\ref{eq:bounds}).
In the context of qubits, we can numerically test that this non-saturation is maximized by a state close to the $\ket{W}$ state.

\subsubsection*{Random tensor example}

Having provided a specific example, we now provide a much more general example that is likely to be relevant to holography.
In the past few years, it has been understood that there are deep connections between random matrix theory and gravity \cite{Hayden:2016cfa,Cotler:2016fpe,Saad:2018bqo,Saad:2019lba,Penington:2019kki}.
In particular, the entanglement structure of holographic states is well approximated by random tensor networks \cite{Hayden:2016cfa}.

\begin{figure}[h]

  \includegraphics[clip,width=0.8\columnwidth]{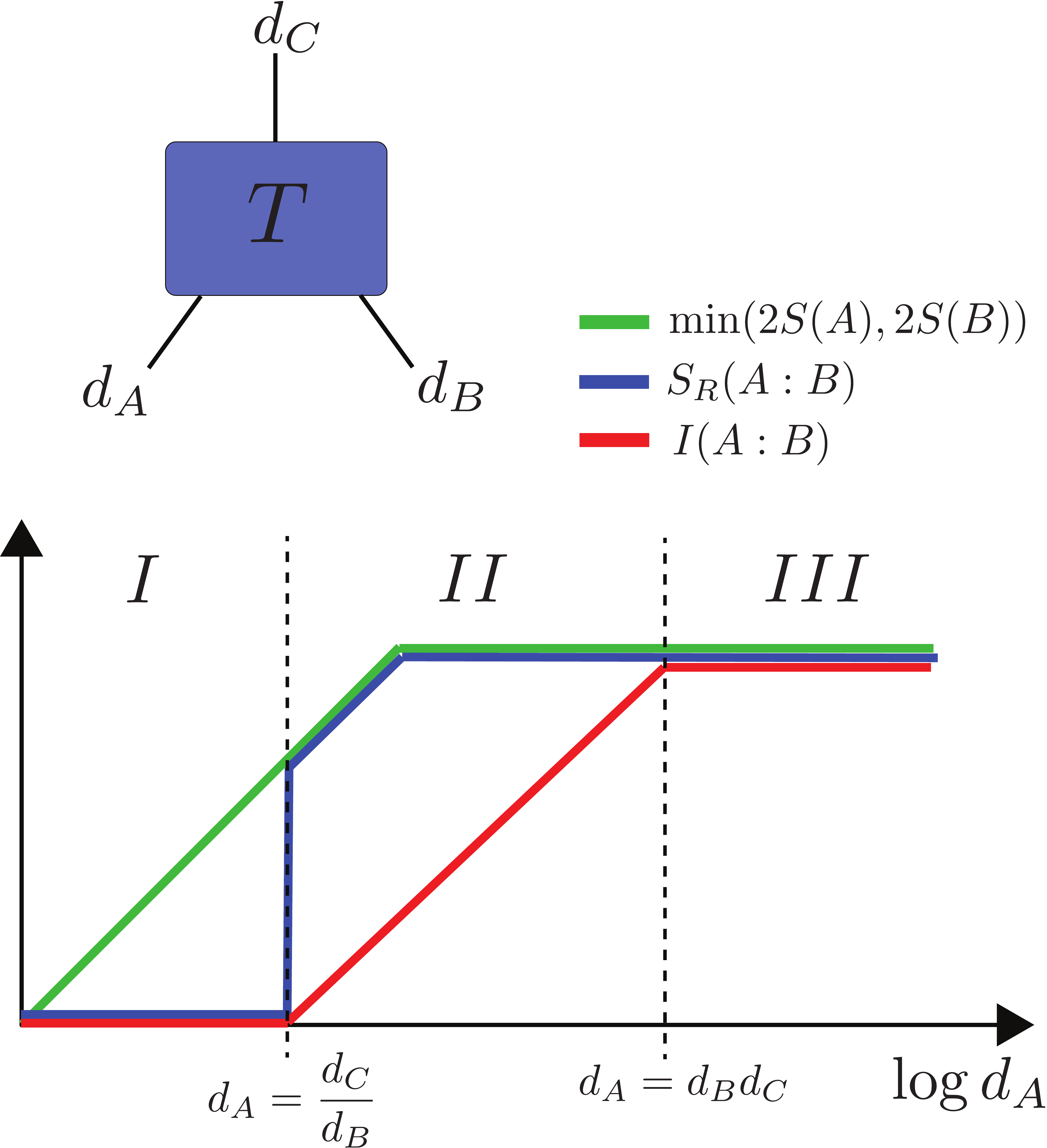}

\caption{A tripartite state comprising a single random tensor $T$ with legs of bond dimensions $d_A$, $d_B$ and $d_C$. We sketch the reflected entropy $S_R(A:B)$ and its upper and lower bounds as a function of the bond dimension $d_A$ while holding $d_B$ and $d_C$ fixed.
}
\label{fig:RTN}
\end{figure}

Motivated by this, we consider the example of a random tripartite state generated by a single random tensor as shown in Figure~\ref{fig:RTN}.
This state is characterized by the dimensions of the individual parties denoted $d_A$, $d_B$ and $d_C$.
This can be thought of as a discrete model of a multiboundary wormhole with three asymptotic boundaries where the bond dimensions are a measure of the area of the mouths of the wormhole \cite{Balasubramanian:2014hda,Marolf:2015vma,Akers:2019nfi}.
This example will be discussed in much more detail in \cite{pagereflected}, but here we use the basic result to emphasize a connection to the results obtained in Section~\ref{sec:examples}.
Essentially, the results described below arise from using the holographic proposals for computing entropy and reflected entropy.

We consider the phase diagram as we hold $d_B$ and $d_C$ fixed such that $d_B^2\geq d_C\geq d_B$, and vary $d_A$.
The large $d$ behaviour of the relevant quantities in this state is sketched in Figure~\ref{fig:RTN} and demonstrates essentially three phases.
Phase I corresponds to the state being dominated by Bell pairs shared separately between $AC$ and $BC$.
Similarly, Phase III corresponds to the state being dominated by Bell pairs shared between $AC$ and $AB$.
Phase II has the interesting feature which is analogous to the $\ket{W_{\epsilon}}$ state discussed in the qubit example where the upper bound is saturated at leading order while the lower bound is far from being saturated.
We can clearly see that this example reproduces precisely the behaviour found in Figure~\ref{fig:plot}.
It would be interesting to use this simple model and probe other correlation measures to make this connection tighter.


\section*{Acknowledgments}
We would like to thank Thomas Faulkner, Raphael Bousso, Yuya Kusuki and Zhenbin Yang for helpful discussions. 
We also thank Tianyi Li, Jinwei Chu and Yang Zhou for cooordinating the release of their paper with us.
This work was supported in part by the Berkeley Center for Theoretical Physics; by the Department of Energy, Office of Science, Office of High Energy Physics under QuantISED Award DE-SC0019380 and under contract DE-AC02-05CH11231; and by the National Science Foundation under grant PHY1820912. 
This work was partly done at KITP during the program “Gravitational Holography”.\\

\appendix
\section{Inequalities of Entanglement of Purification}\label{app:ineq}
The entanglement of purification satisfies an inequality analogous to extensiveness, 
\begin{equation}
S_R(A:B)\leq S_R(A:BC),\label{extensiveness}
\end{equation}
the analogue of which we do not yet know for reflected entropy. 
Let us assume $S_R^{(\text{eff})}$ satisfies this inequality. 
Let us consider the Cauchy slice $\Sigma_{A:B}$ on which $S_R(A:B)$ is optimized.
We can divide the region $D(ABC\cup \text{Is}(ABC))$ into regions $A''$, $B''$ and $C''$ which minimize $S_R(AA'':BB''CC'')_{\Sigma_{A:B}}$ on $\Sigma_{A:B}$. 
On the other hand, by entanglement wedge nesting we know that $D(AB\cup\text{Is}(AB))\subset D(ABC\cup\text{Is}(ABC))$.
Thus, we have $D(AB\cup\text{Is}(AB))\cap\Sigma_{A:B}\subset D(ABC\cup\text{Is}(ABC))\cap\Sigma_{A:B}$.
Given this, we can define $A'=A''\cap D(AB\cup\text{Is}(AB))\cap\Sigma_{A:B}$, and $B'=(B''\cup C'')\cap D(ABC\cup\text{Is}(ABC))\cap\Sigma_{A:B}$.
$A'$ and $B'$ now cover the region of $\Sigma_{A:B}$ inside $D(AB)\cup\text{Is}(AB)$ and can be considered as candidates for the optimization of $S_R(A:B)$.
Using this, we have
\begin{eqnarray}
S_R(A:B)&&\leq S_R(AA'':BB'')_{\Sigma_{A:B}}\leq S_R(AA'':BB''CC'')_{\Sigma_{A:B}}\nonumber\\&&\leq S_R(A:BC),\end{eqnarray}
where the first inequality follows since $S_R(A:B)$ is obtained by minimizing the hybrid entropy on $\Sigma_{A:B}$, the second inequality follows from the assumption that $S_R^{(\text{eff})}$ satisfies the required inequality, and the third inequality follows from maximization.

The entanglement of purification satisfies 
\begin{equation}
I(A:B)+I(A:C)\leq S_R(A:BC),
\end{equation}
which is known to be violated by reflected entropy, at least for classically correlated states.
However, this violation is often invisible in holographic theories, which include a large amount of quantum entanglement. 
We can prove this inequality from our islands formula by assuming that the bulk matter satisfies this inequality, as well as the monogamy of mutual information:
\begin{equation}
I(A:B)+I(A:C)-I(A:BC)\leq 0,\end{equation} which is known to be satisfied by holographic matter. 
Then by using the inequality $S_R(A:B)\geq I(A:B)$, we confirm that this inequality is indeed satisfied.

\bibliography{purification}
\end{document}